\documentclass[%
 reprint,
 amsmath,amssymb,
 aps,
pra,
]{revtex4-2}

\usepackage[toc,page]{appendix}

\usepackage[utf8]{inputenc}
\usepackage{bm}
\usepackage{multirow,amssymb,amsbsy,amsmath}
\usepackage{graphicx}
\usepackage{subfigure}
\usepackage{mathrsfs}
\usepackage{verbatim}
\usepackage{amssymb}
\usepackage{amsfonts}
\usepackage{bbm}
\usepackage{amsmath}
\usepackage{lmodern}
\usepackage{appendix}
\usepackage{soul}
\usepackage[colorlinks,citecolor=blue,urlcolor=blue, linkcolor=blue]{hyperref}
\usepackage{ulem}
\usepackage{xcolor}
\usepackage{dcolumn}
\usepackage{bm}

\newcommand{\bra}[1]{\langle#1|}
\newcommand{\ket}[1]{|#1\rangle}
\newcommand{\hatd}[1]{\hat{#1}^{\dagger}}

\newcommand{\mean}[1]{\langle #1 \rangle}

\newcommand{\proj}[1]{\ket{#1}\!\bra{#1}}

\renewcommand{\aa}[1]{{\color{blue} #1}}

\begin{document}


\title{
Nonclassicality of induced coherence witnessed by 
quantum contextuality
}

\author{F. H. Shafiee}
\thanks{These two authors contributed equally}
\author{O. Mahmoudi}%
\thanks{These two authors contributed equally}
\author{R. Nouroozi}
\author{A. Asadian}
\affiliation{Department of Physics, Institute for Advanced Studies in Basic Sciences (IASBS), Gava Zang, Zanjan 45137-66731, Iran}

\begin{abstract}
Quantum indistinguishability by path identity generates a new way of optical coherence, called ``induced coherence". The phenomenon,  originally uncovered by Zou, Wang, and Mandel's experiment, is an emerging notion in modern quantum experiments with a wide range of implications. However, there has been controversy over its true quantum nature and whether the result can be emulated with classical light. We design a suitable contextuality test that can determine the conditions under which the setting produces distinguishing quantum predictions that cannot be described classically, namely, via the noncontextual hidden variable model.

\end{abstract}

\maketitle


\section{Introduction }
Quantum experiments often challenge our understanding of physical processes. A prominent example is an experiment done by Zou, Wang, and Mandel (ZWM) in 
1991 \cite{ZWM,ZWM2}. Recently, there has been renewed interest in its quantum origin as well as its applications \cite{RMP}. The main innovation of ZWM's work was the introduction of induced coherence without induced emission. Such a brilliant idea can be realized using two similar nonlinear crystals inside a Mach-Zehnder-like interferometer. The schematic representation of the ZWM setup is shown in Fig. \ref {fig1}. These two nonlinear crystals (NL) were pumped by a laser, and spontaneous parametric down-conversion (SPDC) can potentially occur at both, each with a correlated photon pair generation called signal and idler photons. Signal photons ${s}_1$ and ${s}_2$ from NL1 and NL2 respectively, are combined at a beam splitter (BS2) and then collected by detector ${D}_s$. The remarkable point about this approach is that when the ${i}_1$ photon is aligned through NL2 so that two idler photons of ${i}_1$ and ${i}_2$ coincide, then the path identity between them is achieved and the idler photons become indistinguishable, thereby erasing the information about which crystal the signal photon was generated from. As a result of that, the signal photon turns into a superposition of the two modes interfering at the beam splitter (BS2).

In the ZWM experiment, laser light was used to pump the identical nonlinear crystals. The first beam splitter provides equal pumping of both NL crystals. Therefore, when the idler photon from NL1 ($i_1$) is aligned to the idler path of NL2 ($i_2$), the occurrence of stimulated (induced) emission becomes, in principle, unavoidable. But when a weak field of the input pump is used, the average photon number generated by down-conversion is low. This leads to the low amount of generation and amplification of the idler photon in NL2. Their theoretical analysis shows a linear dependence of 
visibility obtained after BS2 against the amplitude transmission coefficient between NL1 and NL2 (O-plate in Fig. \ref{fig1}) that correctly predicted the experimental observations. In this way, it is claimed that the stimulated emission is absent in the concept of induced coherency \cite{ZWM2,ZWM}.

From the practical point of view, ZWM interferometric technique is employed in different applications such as imaging \cite{QuantumimagingZeilinger,Theoryofquantumimaging,example-of-quantum-imaging,classicalimaging,position,perspectives,Interaction-free}, microscopy \cite{microscopy-with-undetected-photons}, spectroscopy \cite{spectroscopy}, generating a light beam in any state of polarization \cite{polarization}, testing the complementarity principle \cite{complementarity,complementarityinbiphoton}, two-color interferometry \cite{interferingphotons}, measuring correlations between two photons \cite{lahiritwin,quantifyingthemomentum}, and generating many-particle entangled states \cite{entanglementbypath,Many-particle,mixed-state-entanglement,multiphotonnon-local}.

From the fundamental point of view, the nonclassicality of induced coherence has been extensively debated over recent years \cite{ZWM,ZWM2,RMP,QuantumimagingZeilinger,lahirinonclassicality,inducedMolmer, controlling,qom,interference-distinguishability,Nature-Review,path-identity-source}.
Wiseman and Molmer compared quantum interference with its classical counterparts and concluded that the induced emission may complete the interference for any finite transmission \cite{inducedMolmer}.
Lahiri $et$ $al.$ showed in a theoretical analysis \cite{lahirinonclassicality} that, for a single-photon pump as a pure quantum state, no emission from the NL1 can stimulate the SPDC at the NL2. {This, however, is challenging to be tested, forasmuch as the statistical behavior of the SPDC gives almost no probability to generate paired photons with single photon pumping. However, the amount of photons in the stimulated emission for low-power pumping (low-power laser as a coherent state) is negligible.}   
In contrast with the above-mentioned references,  Shapiro $et$ $al.$ summarized the concept with a classical or semi-classical interpretation \cite{classicalimaging}.
In another article with a different point of view, Boyd $et$ $al.$ by obtaining an expression for the interference pattern that is valid in both the low- and the high-gain regimes of parametric down-conversion, showed how the coherence of the light emitted by the two crystals can be controlled \cite{controlling}.
{In a follow-up study \cite{ClassicalmodelBoyd} the SPDC process together with the imaging of an object using undetected photons (ZWM setup) were analyzed classically. The reason behind it is the intensity-based measurement in which no nonclassical behavior of detected photons can be extracted.}
This may reflect that merely looking at the interference phenomenon {with simple intensity measurement (first-order correlation)} does not contain enough evidence to rule out a classical explanation, and therefore, one can reproduce the result with classical resources. Thus, reasoning based on interferometric data is convoluted and one needs to look for a more stringent test to identify the quantum nature of the ZWM effect.

Quantum contextuality is a general approach for demonstrating the nonclassical aspect of quantum mechanics \cite{budroni}. Noncontextuality inequalities represent testable constraints imposed by certain classical models. Therefore, the quantum violation of such inequalities indicates the conflict of the quantum predictions with those of classical models or noncontextual hidden variable models in general \cite{kochenandSpecker}.
In the present work, we focus on Kochen-Specker contextuality. Another approach to nonclassicality was proposed by Spekkens which was a generalized notion of contextuality \cite{Spekkens}. The proposed scheme indicates conditions under which the induced coherence by path identity leads to a non-classical result beyond the interference effect. The contextuality tests have the advantage of verifying the quantumess in a ``black-box" scenario. This is because the specifications of the measurements, such as measurement compatibility and sharpness, follow operational definitions that are examined solely from the measurement's outcome statistics without presupposing the validity of quantum mechanics or the inner workings of the devices \cite{budroni}.
 The simulation \footnote[1]{\href{ https://github.com/mahmoudifar/ICWIE}{ https://github.com/mahmoudifar/ICWIE}} of the proposed setup involves the utilization of symbolic algebra \cite{Automated}. (Its details are explained in Appendix \ref{symbol}.)

This article is organized as follows. First, in Section \ref{sec method} the necessary background on the physical model of the ZWM experiment is presented and its basic relations are derived. Then an extended version of the ZWM setup is envisaged to qualify to witness contextuality. In Sec. \ref{sec nc-correlations} an experimentally testable proposal based on the notion of quantum contextuality is introduced and the results are discussed.
\section{Methodology}\label{sec method}

In this section, the key notion of quantum indistinguishability caused by path identity in the ZWM scheme using two nonlinear crystals is briefly reviewed. Afterward, an extended scheme to the three nonlinear crystals is proposed in which illustration of contextual correlations would be possible.

\subsection{ZWM setup: the emergence of induced coherence without induced emission}
\begin{figure}
\includegraphics[width=9cm]{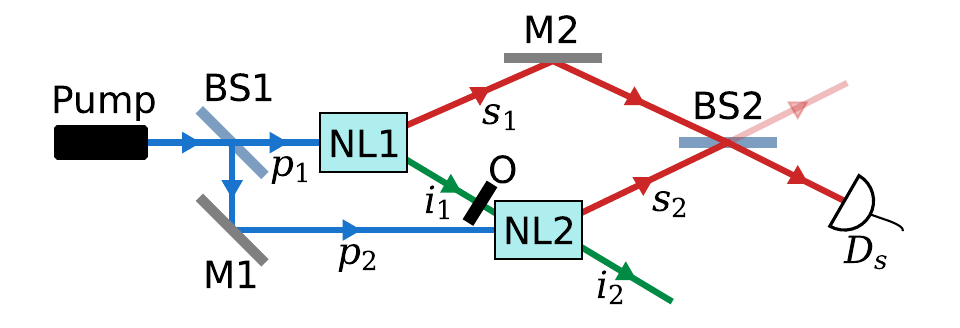}	
\caption{Schematic representation of the ZWM experiment. The identical coherent pump laser can pump two of the same nonlinear crystals, NL1 and NL2, which results in the photon pairs (signal and idler) generation in each crystal. The indistinguishability between idler photons emerges from the perfect alignment of their paths (path identity). Using a second beam splitter BS2, the interference between signal photons appears due to induced coherence yielded by the path identity. Object O with amplitude transmission coefficient $t$ can control the path identity quality.}\label{fig1} 
\end{figure}
The correlated photon pair generation is important for various applications in quantum optics. Besides different methods, the SPDC utilized in the second-order optical nonlinear medium is a promising technique to generate both (heralded) single or correlated paired photons, due to its efficiency and simplicity \cite{SPDC}.
In the SPDC process, a non-linear $\chi^{(2)}$ medium pumped with a single frequency photon of specific energy (${\hbar}\omega_p$) and wave vector ($\vec{k_p}$) to generate two photons of lower energies called signal ($\hbar\omega_s$, $\vec{k_s}$) and idler ($\hbar \omega_i$, $\vec{k_i}$). The process obeys laws of energy ($\hbar\omega_p= \hbar\omega_s+\hbar\omega_i$) and momentum ($\hbar\vec{k_p}=\hbar\vec{k_s}+\hbar\vec{k_i}$) conservations \cite{SPDC,boydbook}. The latter is known as phase matching between interacting waves. The probabilistic behavior of SPDC yields single and multi-pair emissions in low- and high-pump intensities \cite{bulkcrystals}. To distinguish between them, they are referred to as low- and high-gain regimes, respectively.
The Hamiltonian describing SPDC process and the evolution of generated photon pairs,  in the interaction picture, is given by
\begin{equation}
H_{\operatorname{DC}_j}(t)=g e^{i\bigtriangleup\omega t} \hat{a}_{{p}_{j}} \hatd{a}_{{s}_{j}}\hatd{a}_{{i}_{j}} + \operatorname{H.c.}\label{eq1},
\end{equation}
where $j = 1, 2$ labels the nonlinear crystals, $g$ represents the interaction strength, $\bigtriangleup\omega=\omega_s+\omega_i-\omega_p$, $\hat{a}$ and $\hatd{a}$ are photon annihilation and creation operators, respectively. H. c. denotes the Hermitian conjugation \cite{ghosinterference}.
With the unitary operator of SPDC  $\hat{U}_{\operatorname{DC}_j}(t)=\exp{(-iH_{{\operatorname{DC}}_{j}}(t)/\hbar)}$, (see details in Appendix \ref{U for SPDC}) the quantum state becomes
\begin{equation}
\ket{\psi}=\hat{U}_{\operatorname{DC}_j}(t)\ket{\psi_{\rm{in}}},
\label{eq2}
\end{equation}
in which, $\ket{\psi_{\rm in}}$ is the state of light before down-conversion, often being a laser pump coherent state $\ket{\alpha_{p_j}}$ with amplitude $\alpha_{p_j}$ indicating the strength of the pump. Since the first beam splitter causes the same pump for each NL, then $\alpha_{p_1}=\alpha_{p_2}=\alpha_{p}$ (see Fig. \ref{fig1}). Therefore the initial state is
\begin{equation}
\ket{\psi_{\rm{in}}}=U_{\operatorname{BS}_1}\ket{\alpha_p}=\ket{\dfrac{\alpha_{p}}{\sqrt2},\dfrac{\alpha_{p}}{\sqrt2}}.
\label{eq3}
\end{equation}
Based on Eqs. (\ref{eq2}) and (\ref{eq3}), the quantum state of light in the ZWM setup depicted in Fig. \ref{fig1} has the following form after down conversions:
\begin{align}
\ket{\psi}&=\ket{\psi_{\rm{in}}}\otimes\Big(\ket{0_{s_{1,2}},0_{i_{1,2}}}+\dfrac{g\alpha_p}{\sqrt{2}}[\ket{1_{s_1},1_{i_1}}+\ket{ 1_{s_2},1_{i_2}}]\notag\\ 
&+\dfrac{g^{2}\alpha_{p}^{2}}{2}[\ket{2_{s_1},2_{i_1}}+\ket{2_{s_2},2_{i_2}}+\ket{1_{s_1},1_{i_1}}\ket{1_{s_2},1_{i_2}}]\notag\\
&+\dfrac{\sqrt{2} g^{3}\alpha_{p}^{3}}{4}[
\ket{2_{s_1},2_{i_1}}\ket{1_{s_2},1_{i_2}}\notag\\
&+\ket{1_{s_1},1_{i_1}}\ket{2_{s_2},2_{i_2}}]+\cdots\Big).
\label{eq3-1}
\end{align}
The most crucial point in the ZWM scheme is the path identity of two idler photons \cite{ZWM,ZWM2}. This can be achieved by the precise alignment of them in such a way that they can not be distinguished in any degree of freedom. Since during the alignment of two idler photons with the same frequencies ($\omega_{i_1}=\omega_{i_2}$) their polarizations are kept unchanged, perfect alignment means their spatial frequencies must be equal. In the other words, the equal phase matching of the SPDC process in the first crystal is forced to be equivalent to the second one when the perfect spatial frequency matching is achieved. This is reduced to equalizing the wave vector of two idler photons $\vec k_{i_1}=\vec{k_{i_2}}=\vec{k_i}$ which permits to write $\hat{a}_{i_1}(\vec{k_{i_1}})=\hat{a}_{i_2}(\vec{k_{i_2}}) = \hat{a}_{i}(\vec{k_i})$.  The path identity achieved in this way erases the which-crystal information of the final photon in the idler's path. 
The difference between this method and the traditional quantum eraser is that all photons arrive at the same output, regardless of which crystal they are created, rather than being erased by postselection. After the path identity, the state can be written as
\begin{align}
\ket{\psi}&=\ket{\psi_{\rm{in}}}\otimes\Big(\ket{0_{s_{1,2}},0_{i_{1,2}}}+\dfrac{g\alpha_p}{\sqrt{2}}[(\ket{1_{s_1}}+\ket{ 1_{s_2}})\ket{1_i}]\notag\\ 
&+\dfrac{g^{2}\alpha_{p}^{2}}{2}[(\ket{2_{s_1}}+\ket{2_{s_2}}+\sqrt{2}\ket{1_{s_1}}\ket{1_{s_2}})\ket{2_{i}}]\notag\\
&+\dfrac{\sqrt{6} g^{3}\alpha_{p}^{3}}{4}[(\ket{2_{s_1}}\ket{1_{s_2}}+\ket{1_{s_1}}\ket{2_{s_2}})\ket{ 3_i}]+\cdots\Big).
\label{eq4}
\end{align}
In Eq. (\ref{eq4}), two interaction regimes of SPDC are evident. In the low-gain regime, the input laser light is so weak that the simultaneous generation of single photon-pair in both crystals is negligible. In addition, the probability of multiphoton-pair generation is low, as well. Therefore in such a regime of interaction, only one photon of each signal and idler is in the setup. This leads to the absence of the necessary idler photon for the stimulation of induced emission.
This regime of interaction which is related to the term proportional to $\alpha_{p}$ in Eq. (\ref{eq4}) acts as the single photon pumped regime. Under this condition, Eq. (\ref{eq4}) is reduced to
\begin{equation}
\ket{\psi}\approx\ket{\psi_{\rm{in}}}\otimes\Big(\ket{0_{s_{1,2}}, 0_{i_{1,2}}}+\dfrac{g\alpha_p}{\sqrt{2}}[(\ket{1_{s_1}}+\ket{ 1_{s_2}})\ket{1_i}]\Big).
\label{eq5}
\end{equation}
Another interaction regime in Eq. (\ref{eq4}) is related to the high-gain regime (the term $\alpha_{p}^{2}$  and higher order of $\alpha_{p}$). For such a strong pumping the first beamsplitter provides sufficient pumps for both NLs. Therefore, the photon-pair generation occurs in each NL. Here two types of stimulated processes may be accessible: (i) induced emission due to stimulation of SPDC process by multi-pair generation in high pump power inside of each NL ($\ket{2_{s_1}}$ and $\ket{2_{s_2}}$); (ii) induced emission by idler photons that their path identity is achieved ($\sqrt{2}\ket{1_{s_1}}\ket{1_{s_2}}$). In the higher order of $\alpha_{p}$, however, these two types of processes appear simultaneously ($\ket{2_{s_1}}\ket{1_{s_2}}$ and $\ket{1_{s_1}}\ket{2_{s_2}}$). Similar to \cite{controlling}, the term $\alpha_{p}^{2}$ is known as a high-gain source and the higher order of $\alpha_{p}$s refers to the high-gain regime. Again, it is worth mentioning that, in the low-gain regime, induced coherence is most probable, while induced emission dominates in the high-gain source and regime.

Now, to observe an unambiguous quantum mechanical distinction, nonclassical correlation measurements of generated photons are needed. This can be done using quantum contextuality as a general framework of a nonclassicality test in which a simple noncontextual inequality should be tested. To examine such a method, the extension of the ZWM setup from two to at least three crystals is necessary. This, including the results of different pumping regimes, is presented in the next subsection.

\subsection{Extended version of ZWM scheme }
\begin{figure}
\raggedright
\subfigure[]{\label{2aa}}
\includegraphics[width=8.6cm]{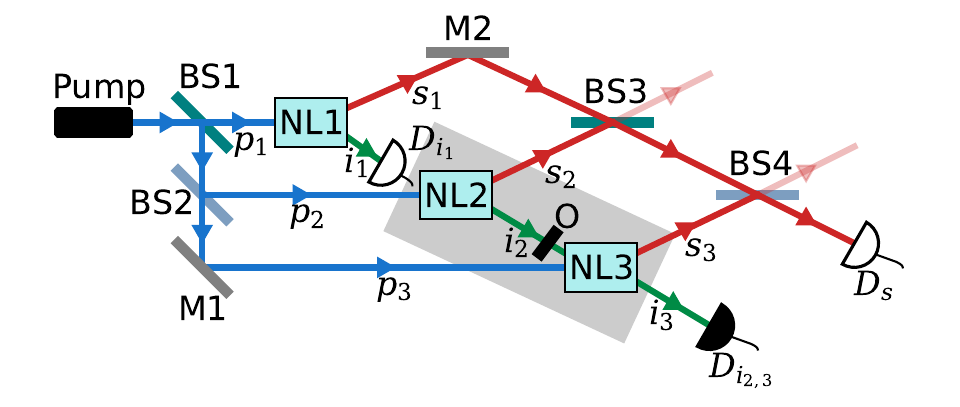}
\raggedright
\subfigure[]{\label{2b}}
\includegraphics[width=8.6cm]{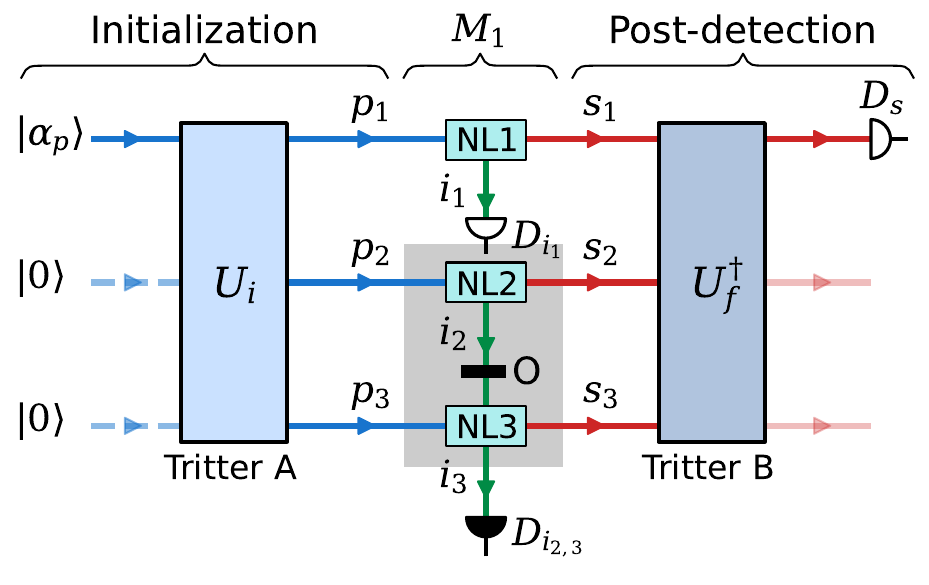}
\caption{(a) Extended version of the ZWM experiment. Three similar nonlinear crystals (NL1, NL2, and NL3) can be pumped with the same coherent pump laser. Each crystal can potentially emit bi-photons via spontaneous parametric down-conversion (the idlers $\mathrm{i_1}$, $\mathrm{i_2}$, and $\mathrm{i_3}$ and the signals $\mathrm{s_1}$, $\mathrm{s_2}$, and $\mathrm{s_3}$). Signal photons are combined by two different beam splitters $\mathrm{BS3}$ and $\mathrm{BS4}$, then counted with the detector $D_s$. The gray rectangle shows the path identity between the idler $\mathrm i_2$ and $\mathrm i_3$ in one measurement.
(b) Conceptual circuit-like representation. The first tritter causes the crystals to be pumped equally, and the second tritter performs transformation for a specific postselection state. Detectors can either be on (white color) or off (black color).
 }\label{fig2} 
\end{figure}

In the original ZWM setup shown in Fig. \ref{fig1}, the operations and measurements of interest are defined in two-dimensional Hilbert space. Since the proposed scheme studied here is based on contextuality, the operational dimension should be increased. According to the Kochen-Specker theorem, the minimal dimension to witness contextuality is $d=3$ \cite{kochenandSpecker}. To implement such tests for a ZWM-based concept, extended three-path interferometry is proposed. The new scheme which is shown both schematically and operationally in Figs. \ref{2aa} and \ref{2b},  respectively, consists of a straightforward extension of the ZWM scheme to three paths in which the path identity can be applied consecutively in two of these three paths (NL1 and NL2 or NL2 and NL3) for each set of measurement explained below.
The three nonlinear crystals can be pumped equally to generate a pair of correlated signal and idler photons. Tritter ${A}$ is the combination of two beam splitters and a mirror and prepares the same pump for each NL. ($\operatorname{BS1}$ and $\operatorname{BS2}$ with transmission coefficients of $\tfrac{1}{\sqrt{3}}$ and $\tfrac{1}{\sqrt{2}}$, respectively, together with mirror $\mathrm{M1}$ in Fig. 
 \ref{2aa} generates $U_i$ in Fig. \ref{2b} \cite{multiportBS,multiphotonBS}). Therefore,  the laser pump, described by a coherent state with amplitude $\alpha_p$, under the first tritter $U_i$ splits into equally weighted beams with the amplitude of $\frac{\alpha_p}{\sqrt{3}}$. Then the input state of the scheme can be represented as:
 \begin{equation}
 \ket{\psi_{\rm{in}}}=\ U_{i} \ket{\alpha_{p}, 0, 0}=\ket{\frac{\alpha_p}{\sqrt{3}},\frac{\alpha_p}{\sqrt{3}},\frac{ \alpha_p}{\sqrt{3}}}.
 \label{eq6}
\end{equation}
Each beam illuminates a nonlinear crystal in which the SPDC process can generate photon pairs as follows:
\begin{align}
\ket{\psi}&= U_{\operatorname{DC}_3}U_{\operatorname{DC}_2}U_{\operatorname{DC}_1}\ket{\psi_{in}}=\notag
\ket{\psi_{in}}\otimes\Big(\ket{0_{s_{1,2,3}},0_{i_{1,2,3}}}\notag\\
&+\frac{g\alpha_p}{\sqrt3}[\ket{1_{s_1}, 1_{i_1}}+\ket{1_{s_2}, 1_{i_2}}+\ket{1_{s_3}, 1_{i_3}}]+\cdots\Big).
\label{eq7}
\end{align}
Note that $\alpha_p$ is equal for each NL. In addition, different pumping regimes, similar to the two-crystal scheme, can be characterized here, as well. In fact, this is essential while it provides useful tools to compare results from the original ZWM setup with two crystals with the extended version of that with three crystals.

The key feature in the extended version of the ZWM scheme is the indistinguishability of successive generated idler photons either in NL1 and NL2 ($i_1$ = $ i_2$ = $i$) or NL2 and NL3 ($ i_2$ = $ i_3$ = $ i$) when their path identity is satisfied. Since these crystals are usually pumped by laser beams, there is always a nonzero probability of the presence of idler photons generated by NL1 at the NL2 or NL2 at the NL3, when the down-conversion is taking place at the latter. Therefore, the stimulated (induced) emission at NL2 or NL3 occurs.
To vary the path identity of generated idler photons, one can insert a beam splitter between two  NLs (NL1 and NL2 or NL2 and NL3). In this case (when placed between NL2 and NL3), 
 $i_2$ and $ i_3$ are connected by
\begin{equation}
\hat{a}_{i_3}=t\hat{a}_{i_2} +r\hat{a}_0
\label{eq8},
\end{equation}
where $t$ as transmissivity and $r$ as reflectivity amplitudes of beamsplitter obey the $ \left|t \right|^{2} +\left|r\right|^{2}=\left|T\right|+\left|R \right|=1$ relationship. $\hat{a}_{0}$ describes the vacuum field at the unused port of the beam splitter. This leads to the evaluated state of the low-gain regime as
\begin{align}
\ket{\psi}\approx&\ket{\psi_{\rm{in}}}\otimes\Big(\ket{0_{s_{1,2,3}},0_{i_{1,2,3}}}+\frac{g\alpha_p}{\sqrt{3}}[\ket{1_{s_1}}\ket{1_{i_1}}\notag\\
+&(t\ket{1_{s_2}}+\ket{1_{s_3}})\ket{1_i} +r\ket{1_{s_2}}\ket{1_0}]\Big)\label{eq9}. 
\end{align}
In the following, a detailed discussion of the expected outcome of the proposed three-crystal scheme is presented in detail.
{Along with previous works \cite{lahirinonclassicality, controlling, ClassicalmodelBoyd}, the visibility and coincident detection rate of signal photons are investigated in different pumping regimes, and their limitation for providing conclusive evidence of nonclassicality is outlined.} 
\subsubsection{Visibility}
In previous works \cite{ZWM,ZWM2,RMP,lahirinonclassicality,lahirican,interference-distinguishability,inducedMolmer,controlling,qom,Nature-Review,complementarity,complementarityinbiphoton}, the main approach addressing the quantum behavior of ZWM was based on the visibility of the interference.
The photon counting rate using the annihilation and creation operators after the second tritter $U_f$ (Similar to the $U_i$, the $U_f$ is the combination of two beam splitters $\mathrm{BS3}$ and $\mathrm{BS4}$ with additional $\mathrm\pi$ phase shift, respectively. see details in Appendix \ref{tritter}) at the detector $D_s$ is given by: 

\begin{equation}
R_s = \mean{\psi|\hatd{a}_{s}\hat{a}_{s}|\psi}.
\label{eq10}
\end{equation}

Thereby, the resultant photon counting rate for the single photon pumping (low-gain regime) after the second tritter using state Eq. (\ref{eq9}) becomes,
\begin{equation}
R_s=\frac{g^2\alpha^2_p}{3} (1+\dfrac{{t}}{\sqrt{3}}\cos\phi_{s})
\label{eq11},
\end{equation}
where $\phi_{s}$ is the relative phase due to the different optical paths of signal photons from NL2 to NL3.
From Eq. (\ref{eq11}), the visibility of interfering signal photons can be obtained as
\begin{equation}
\nu=\frac{R_{s\max }-R_{s\min }}{R_{s\max }+R_{s\min }}.\label{eq12}
\end{equation}

\begin{figure}
\includegraphics[width=8.6cm]{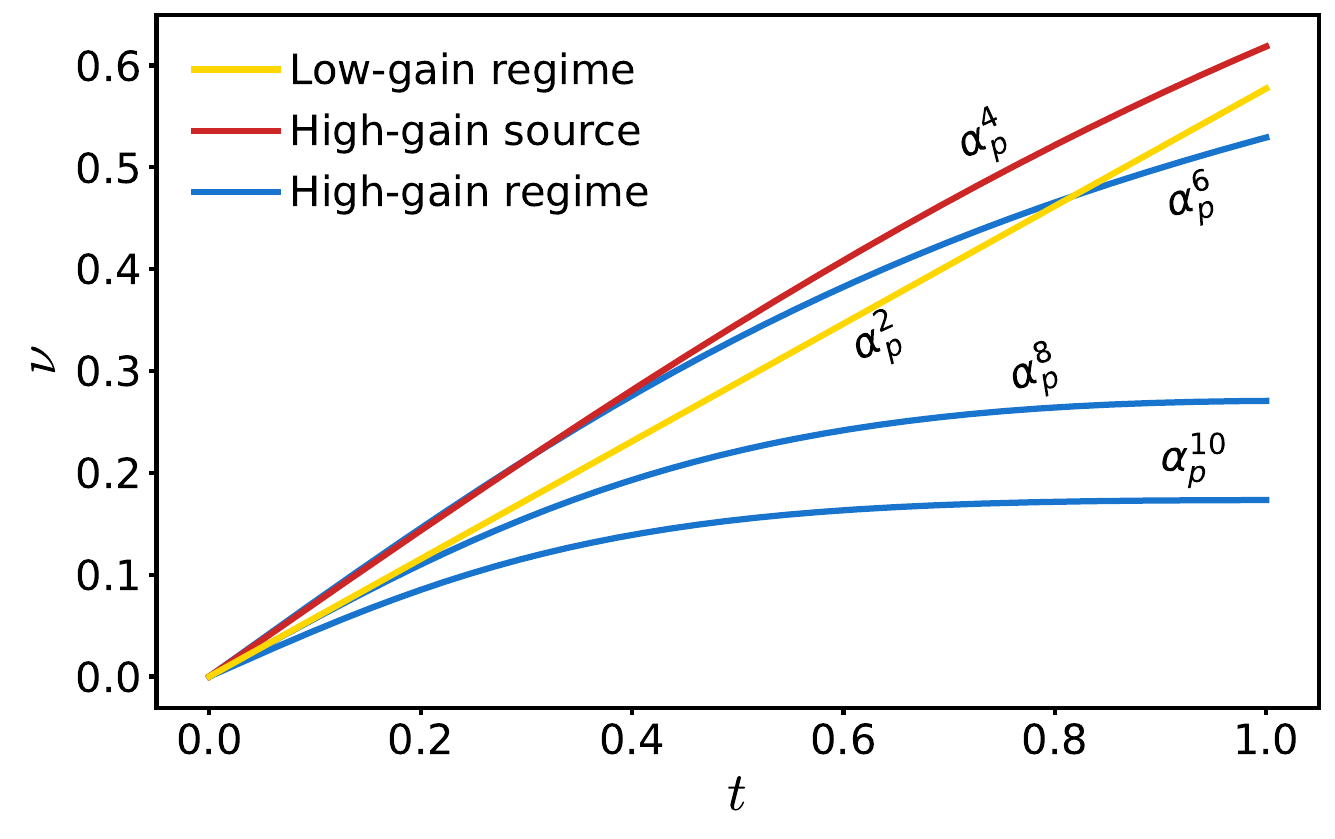}
	\caption{The visibility as a function of path identity variation (controlled by beam splitter transmissivity $t$) for different regimes of operation. Only the yellow curve $\alpha_{p}^{2}$, which is for the low-gain regime (roughly the single-photon pumping) has a linear relationship with ${t}$. The visibility for the high-gain source (red curve $\alpha_{p}^{4}$) is larger than the high-gain regime (blue curve $\alpha_{p}^{6}$, $\alpha_{p}^{8}$, and $\alpha_{p}^{10}$). For higher transmissivity, the visibility increases due to the path identity of idler photons.}\label{fig3}
\end{figure}
\begin{figure}
\includegraphics[width=8.6cm]{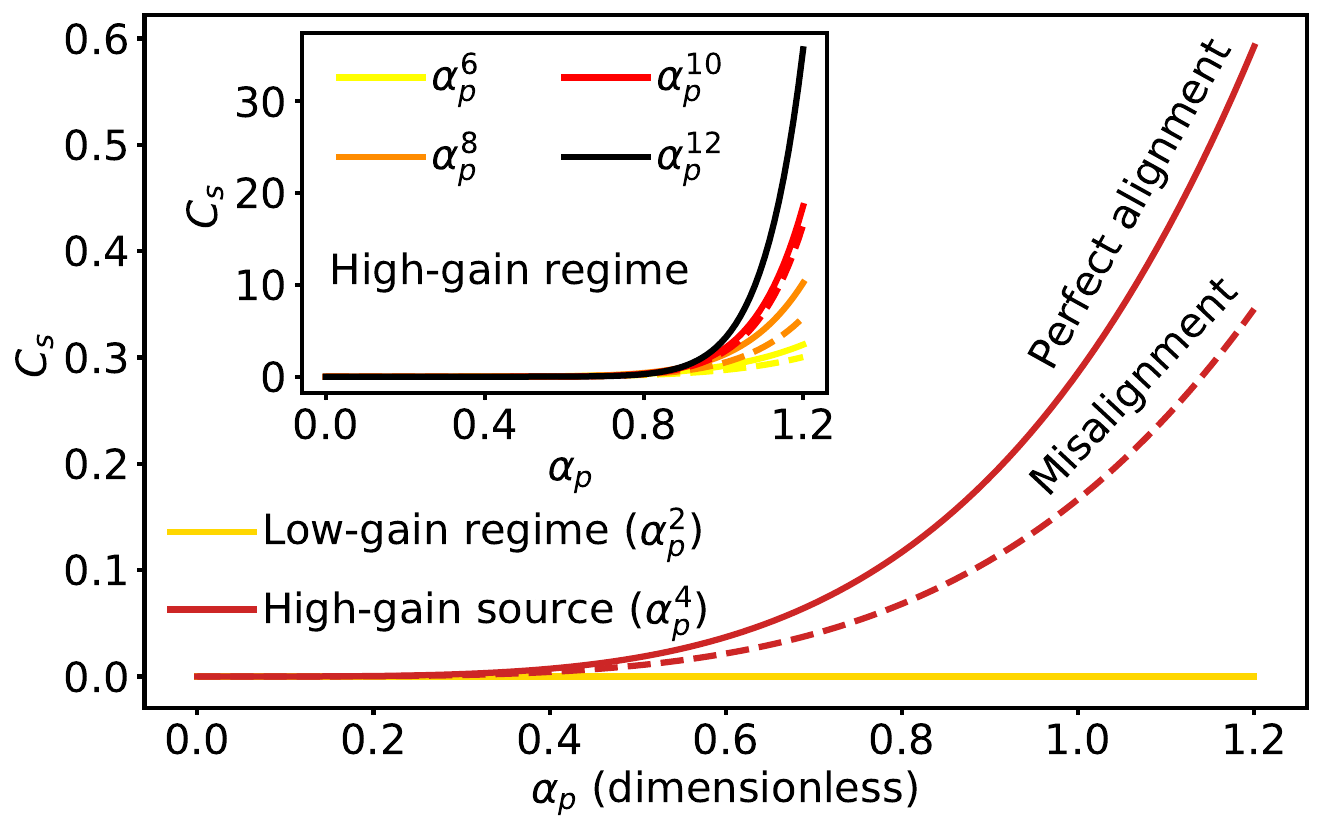}
\caption{(Color online) The coincidence detection rate (dimensionless unit) for different regimes. The full curves show the idler photons are fully aligned for NL2 and NL3 and the dashed curves show the idler photons are fully misaligned.  The yellow curve (light gray) shows the low-gain regime and the red one (dark gray) shows the high-gain source. In the inside subplot, the high-gain regime for different orders of $\alpha_p$ is plotted. The yellow (light gray), orange (medium gray), red (dark gray), and black (black) curves show $\alpha_{p}^{6}$, $\alpha_{p}^{8}$, $\alpha_{p}^{10}$ and $\alpha_{p}^{12}$, respectively. In all cases of the high-gain regime, the coincidence detection rate increases with the growth of the pump power. However, if the idlers are fully misaligned, due to loss of path identity the coincidence rate becomes smaller than the value obtained for fully aligned idlers.}\label{fig4} 
\end{figure}
As expected for the low-gain regime, visibility is linearly proportional to the modulus of the amplitude transmission coefficient $t$ {[the yellow curve ($\alpha_{p}^{2}$) in Fig. \ref{fig3}]}. The physical reason for the non-classicality of linear dependence is that stimulated emission is suppressed, and only spontaneous emission occurs. As mentioned before, since pumping nonlinear crystals usually is via laser light, the induced emission may occur in the experiment. Therefore, for high gain, visibility no longer has a linear dependence on $t$ {the blue curves ($\alpha_{p}^{6}$, $\alpha_{p}^{8}$ and $\alpha_{p}^{10}$) in Fig. \ref{fig3}}.
The visibility for the high-gain source is larger than the low-gain regime {the red curve ($\alpha_{p}^{4}$) in Fig. \ref{fig3}}. There is an intuitive explanation for this fact: due to the seeding of NL3 with generated idler by NL2, induced emission causes more photon generation in NL3 in comparison with NL2. Hence, both arms become unevenly populated, and with raised amplitude transmissivity, the visibility increases. In high-gain regimes, visibility decreases remarkably. This is again due to more induced emission which results in the reduction of induced coherence. In comparison between these three regimes of interaction, one can conclude that visibility follows induced coherency. In addition, with increasing transmissivity, increasing in visibility is observed. This is a direct consequence of path identity influence. In general, the path identity together with its consequence of induced coherency enhances the visibility which is in contrast to induced emission that causes decreasing in visibility (Fig. \ref{fig3}).
\subsubsection{Coincidence detection rate}

To view the appearance of induced coherency without induced emission in different regimes, the coincidence detection rate can be calculated  \cite{lahirinonclassicality,Theoryofquantumimaging,controlling}. The coincidence detection rate after down-converters is given by
\begin{equation}
C_{s}=\mean{\psi|\hatd{a}_{s_2}\hatd{a}_{s_3}\hat{a}_{s_3}\hat{a}_{s_2}|\psi}\label{eq13}.
\end{equation}

Figure. \ref{fig4},  shows the coincidence detection rate $C_{s}$ for different regimes against the pump power when the idler photons are fully aligned (full curves) and misaligned (dashed curves) between NL2 and NL3. In the case of the low-gain regime (single photon regime) no stimulated emission is present, even with fully aligned idler photons. Therefore, the coincidence detection rate remains unchanged (zero) when the idler photons are misaligned. In the high-gain regime, the alignment of the idler photons causes the induced emission, and the curve of fully aligned shows a higher counting rate compared with the fully misaligned one.

{Results presented in this section claim that, a challenge to distinguish between induced coherency and emission still remains. The visibility obtained from the low-gain regime and high-gain source is roughly the same, and the coincident detection of the low-gain regime and the high-gain source is as well. In addition, distinguishing between perfect alignment (with path identity) and misaligned (no path identity) idlers in high-gain sources is problematic with coincident detection. Moreover and most importantly, as it is claimed in \cite{ClassicalmodelBoyd, clHOM}, both the visibility and the coincident detection rate are intensity-based measurements in which the nonclassicality of the detected light can not be ensured.}

{Induced coherence without induced emission (to a good approximation) occurs at low-gain input, but coincidence detection alone cannot confirm it. A feasible nonclassicality witness which goes beyond measuring visibility and coincidence detection is therefore necessary. In the next section contextuality based on the Kochen-Specker theorem is introduced for the extended version of ZWM setup with three NLs. The setup is capable of performing genuine sequential measurements on photons. That is, the joint probability of the two outcomes is read out directly from two detectors, coincidence detection, or in other words evaluating second-order coherence. The necessity of sequential implementation of the joint measurement, in which the individual observables of a given context (correlated measurements) are measured directly and used again in other contexts, is discussed in detail in a recent review in \cite{budroni} as a formal requirement of the Kochen-Specker contextuality test. This requirement in any test of contextuality with photons was also emphasized in \cite{BIC}. The shortcoming of previous implementations is that the joint probabilities are estimated from local intensities \cite{experimentalHardy-likequantum}, characterized by first-order coherence. Consequently, any photonic test whose measurements constitute only 
first-order coherence can be simulated with the classical theory of coherence
\cite{BIC}.
Our setup provides an improved test of nonclassicality based on sequential measurements measured with second-order coherence. Thus, only the induced coherence (with nonclassical property) can pass it.}

\section{nonclassical correlations generated by path identity} \label{sec nc-correlations}

In this section, the correlation measurements probing the nonclassical aspect of ZWM based on the notion of Kochen-Specker (KS) contextuality are developed. It is started by illustrating how the proposed setup, realizes an interesting example of contextual behavior known as Hardy-like contextuality \cite{Hardy-like}. Then the description of its implementation in the proposed setup is presented.

In the following, it is shown that the extended ZWM setup assisted by path identity fulfills the required conditions for implementing joint measurements of compatible observables. 

Note that the photon generation in SPDC process is probabilistic and therefore the registered experiment for estimating the outcomes' probabilities is identified by projection onto a non-vacuum state. That is, a laser pump that fails to produce photons, in which case none of the detectors click, should be discarded.

After the full alignment of idler beams ($i_1= i_2=  i_3= i$), the state of the idler and signal modes in the second line in Eq. (\ref{eq7}) reduces to
\begin{align}
\ket{\psi_{is}}=\frac{1}{\sqrt3}\Big(\ket{1_{s_1}}+\ket{1_{s_2}}+\ket{1_{s_3}}\Big)\ket{1_i},
\label{eq17}
\end{align}
in the single-photon limit. 
This means that the single photon being in signal modes turns into a quantum coherent superposition of the three signal paths due to the indistinguishability of idler paths, which is known as path identity. Here, $\ket{\psi_s}=\frac{1}{\sqrt3}(\ket{1_{s_1}}+\ket{1_{s_2}}+\ket{1_{s_3}})$ constitute a desired  pre-selected state. 

We use so-called on-off detectors to perform dichotomic measurements. For example, if we place the detector in idler beam $i_1$ the corresponding projectors are
 \begin{align}
    \Pi_{i_1}^{\rm on}= \mathbbm 1_{i_1}-\Pi_{i_1}^{{\rm off}} \ \ \ , \ \ \  \Pi_{i_1}^{{\rm off}}=\proj{{\rm vac}_{i_1}}.
    \label{eq19}
\end{align}
Of course, in the single-photon regime, the detector projectors effectively reduce to $\Pi^{\rm on}_{i_1}\equiv\proj{1_{i_1}}$.

A particular postselection (second measurement) is considered a dichotomic measurement performed on the signal photon with the associate projector being
\begin{equation}
    \Pi_{\rm post}^{\rm on}=U_f(\mathbbm 1-\proj{{\rm vac_{s_1}}})U_f^\dag.
    \label{eq18}
\end{equation}
In the single photon limit $\Pi_{\rm post}^{\rm on}=U_f \proj{1_{s_1}} U^\dag_f$
and  the failure detection is described by $\Pi_{\rm post}^{\rm off}= U_f\proj{{\rm vac}_{s_1}}U_f^\dag$. The unitary $U_f$ is defined with effect, 
\begin{equation}
 \ket{\psi_{\rm post}}=U_f\ket{1_{s_1}}=(\ket{1_{s_1}}-\ket{1_{s_2}}+\ket{1_{s_3}})/\sqrt{3}   
\end{equation} 

The key fact about our model is that the idler modes serve as an ancillary read-out system for sensing signal photons indirectly. This enables us to perform sequential‌ measurements. The first measurement is performed on idler photons and the second measurement on signal photons. The photon detection of the detector $D_{i_1}$ is described by $\Pi^{ \rm on}_{i_1}$ and the detector $D_{i_{2,3}}$ is described by $\Pi^{ \rm on}_{i_{2,3}}$, which does not resolve as to which crystal NL2 or NL3 the photon was generated from. This intrinsic loss of information turns the signal photon into a superposition between $ s_2$ and $ s_3$.  Therefore, before the idler photon is detected the faithful form of $\ket{\psi_{s,i}}$ due to the detector locations should read as
\begin{equation}
  \frac{1}{\sqrt3}\ket{1_{s_1}} \ket{1_i}+\sqrt \frac{2}{3}\Big(\frac{\ket{1_{s_2}}+\ket{1_{s_3}}}{\sqrt 2}\Big)\ket{1_{i_{2,3}}}.
\label{eq20}
\end{equation}

The measurement scheme performed in the extended ZWM setup is capable of implementing a testable noncontextual inequality explained in the following section.


\subsection{Hardy-like quantum contextuality}
\label{subsec Hardy-like}
\begin{figure}
\includegraphics[width=9.2cm]{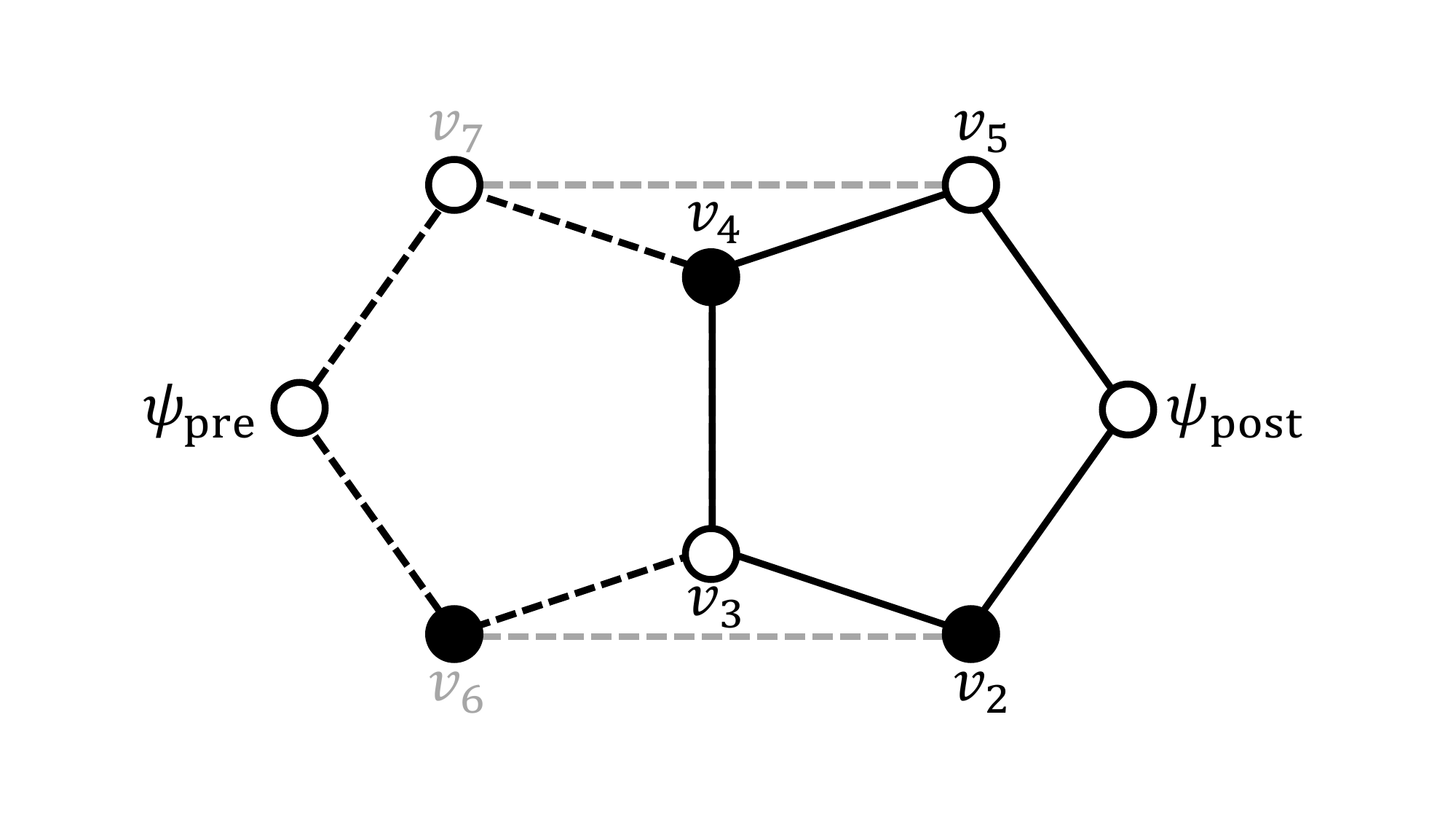}
    \caption{The orthogonality relationships between the five mutually orthogonal vectors (events) help to present them in a pentagon-shaped Kochen-Specker subgraph. The vertices represent measurable events and the edges represent the exclusivity relation between the events.  In the NCHV theory the logical values are assigned to all events, i.e., all vertices need to be colored either black or white. The white vertices show the detected events and the black vertices represent the undetected ones. The graph structure implies the paradox.}
    \label{fig5}
\end{figure}

{The link between pre- and postselection paradoxes and Hardy nonlocality argument are known \cite{PPhardy}, and in a sense, all Hardy-like arguments can be translated as Hardy-like contextuality arguments \cite{private}.} 
The Hardy-like paradox presents a simple proof of quantum contextuality \cite{Hardy-like,experimentalHardy-likequantum} that follows a similar argument to the three-box paradox in the sense that both are examples of logical contextuality. Imagine five boxes, numbered from 1 to 5, where each can be either empty $(0)$ or full $(1)$. Let us denote $P(0,1 \arrowvert 1,2)$ as the joint probability of finding box 1 empty and box 2 full. One can construct a system in a state such that
\begin{subequations} \label{subeqs1}
	\begin{equation}
	P(0,1|1,2) + P(0,1|2, 3)=1,\label{22a}
	\end{equation}
	\begin{equation}
	P(0,1|3, 4) + P(0,1|4, 5)=1.
	\end{equation}
\end{subequations}
Condition Eq. (\ref{22a}) means that when box 1 is empty then box 2 is full and when box 2 is empty then box 3 is full. The above conditions imply
\begin{equation}
P(0,1|5,1)=0,
\label{eq23}
\end{equation}
This is true for the case in which opening the boxes reveals the predetermined values. We can depict these conditions by coloring the vertices of a pentagram with white or black color indicating mutually exclusive events, see Fig. \ref{fig5}.

In quantum mechanics, however, one can prepare a quantum system in state $\ket{\psi_{\rm pre}}$ and five different projections onto,
\begin{subequations} \label{subeqs2}
    \begin{align}
       \ket{v_1}&=\ket{\psi_{\rm post}},\label{v1} \\
        \ket{v_2}&=\frac{1}{\sqrt{2}}\big(\ket{1}+\ket{2}\big), \label{v2}\\
        \ket{v_3}&=\ket{3}, \label{v3}   \\
        \ket{v_4}&=\ket{1}, \label{v4} \\
        \ket{v_5}&=\frac{1}{\sqrt{2}} \big(\ket{2}+\ket{3}\big).
    \end{align}
\end{subequations}
which contradicts the classical result \eqref{eq23}.

In the following, we wish to realize a correspondence between the above projectors and the detection scheme employed in the extended ZWM setup Fig. \ref{fig2}. We translate the above projectors into the corresponding detector's projectors as follows:  $\proj{v_1} \rightarrow \Pi^{\rm on}_{\rm post}$,  $\proj{v_2} \rightarrow \Pi^{\rm on}_{i_{1,2}}$, $\proj{v_3}\rightarrow \Pi^{\rm on}_{s_3}$, $\proj{v_4} \rightarrow \Pi^{\rm on}_{i_1}$, and $\proj{v_5} \rightarrow \Pi^{\rm on}_{i_{2,3}}$. Note that $\ket{\psi_{\rm pre}}\equiv \ket{\psi_{s,i}}$. 

Therefore, this identification enables us to implement the joint probabilities in Eqs. (\ref{subeqs1}) and (\ref{eq23}) in our setup. For example,
\begin{equation}
    P( 0, 1 | 1, 2)= \mean{\Pi^{\rm off}_{\rm post} \Pi^{\rm on}_{i_{1,2}}}.
     \label{eq25}
\end{equation}
Indicating the joint probability of detecting photons at detector $D_{{i}_{1,2}}$ while no photon detection occurs at detector $D_s$. The other joint probability which explicitly indicates the contradiction between quantum and classical prediction is
\begin{equation}
     P(0, 1| 5, 1)= \mean{\Pi^{\rm off}_{i_{2,3}} \Pi^{\rm on}_{\rm post}}=\frac{1}{9},
     \label{eq26}
\end{equation}
predicting a non-zero probability that measuring  $1$ and $5$ reveal different values. 
Kochen and Specker's graph, Fig. \ref{fig5} demonstrates the contradiction between QT and NCHV theories by a set of eight vectors \cite{kochenandSpecker,kochen,budroni,two-KS,kS-for-eight-dimensional,experimental-Ks}. 
 The five vectors in Eqs. (\ref{subeqs2}) are represented by a pentagon-shaped Kochen-Specker subgraph. These vectors lead to the maximum quantum violation of the KCBS inequality \cite{KCBSbound,Hardysnonlocality,cabello,basic}.

\subsection{Results for witnessing contextuality }
\label{subsec KCBS}
Hardy-like proof of contextuality is a particular violation of the simplest noncontextual inequality connected to the Kochen-Specker theorem \cite{cabello-KS,Hardy-like,simpleKCBS}. Based on that theorem, the Klyachko, Can, Binicio\u glu, and Shumovsky (KCBS) inequality and its maximum quantum bound \cite{KCBSbound,simpleKCBS,mathematicalbook} can be written as
\begin{equation}
\kappa=\sum_{i=1}^{i=5} P(0,1 |i, i+1) \overset{\rm NCHV}{\le} 2 \overset{Q}{\le} \sqrt{5}. \label{eq27}
\end{equation}
The left side of Eq. \eqref{eq27} is the sum of the five joint probabilities introduced above. For NCHV theories \cite{cabello-noncontext,kCBS,KCBSbound}, this sum is upper bounded by 2.

Performing the above inequality by our detection scheme is explicitly expressed as
\begin{align}
    \kappa=\mean{ \Pi^{\rm off}_{\rm post} \Pi^{\rm on}_{i_{1,2}}}+&\mean{ \Pi^{\rm off}_{i_{1,2}} \Pi^{\rm on}_{s_3}} + \mean{ \Pi^{\rm off}_{s_3} \Pi^{\rm on}_{i_1} } \\ \nonumber
    + &\mean{ \Pi^{\rm off}_{i_1}\Pi^{\rm on}_{i_{2,3}} }+ \mean{ \Pi^{\rm off}_{i_{2,3}} \Pi^{\rm on}_{\rm post} } \leq 2.
    \label{eq28}
\end{align}
For instance, the placement of detectors in Fig. \ref{fig2} represents the implementation of $P(0,1 |5,1)=\mean{ \Pi^{\rm off}_{i_{2,3}} \Pi^{\rm on}_{\rm post} }$.
Our correlation detections fulfill Eqs. \eqref{subeqs1} and \eqref{eq26} in the low-gain regime where the single-photon generation is more likely to occur. However, Eq. \eqref{eq26} is obtained contradicting the classical prediction, giving $2+1/9$. Therefore, the result is a particular violation of the KCBS inequality. 
\begin{figure}
\includegraphics[width=8.6cm]{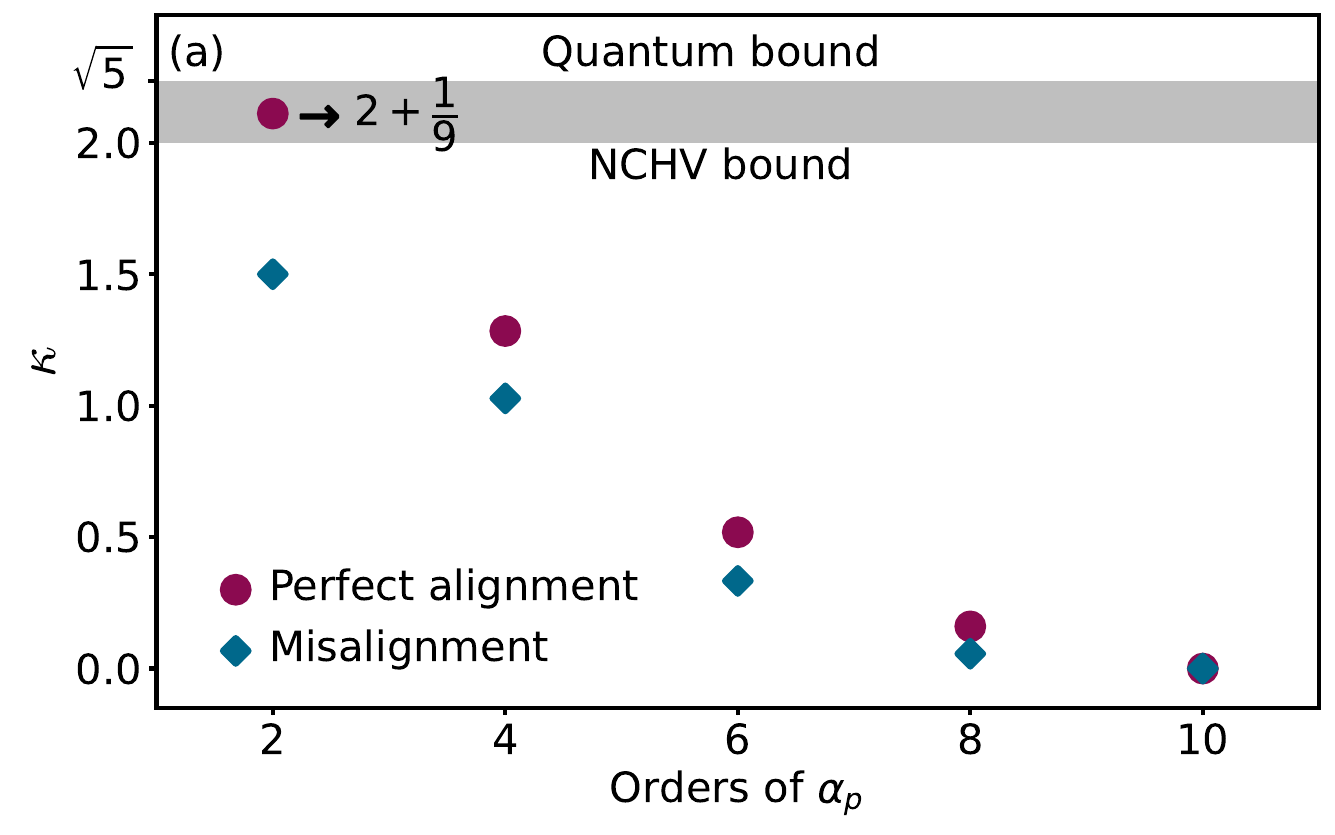}
\includegraphics[width=8.6cm]
{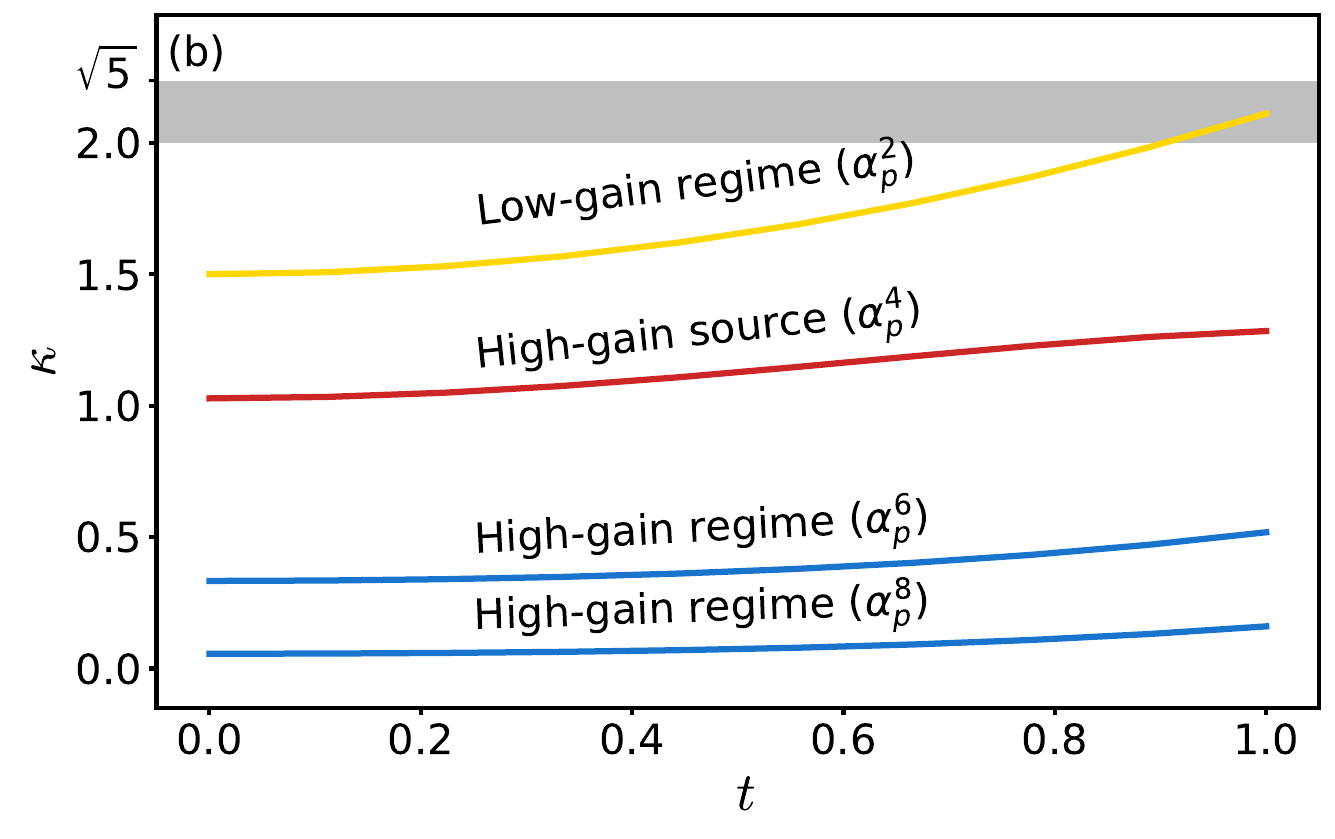}
	\caption{ (a) Results for KCBS inequality shown for different regimes of the laser pump. $\kappa$ is obtained for the two cases when the idler beams are perfectly aligned and the beams are completely misaligned.
 (b) KCBS inequality when the beam splitters between nonlinear crystals control the quality of path identity. As the amplitude transmissivity $t$ increases, the path identity becomes complete, leading to violated KCBS inequality for the low-gain regime.}
 \label{fig6}
\end{figure}
Note that the maximum violation is achieved in the low gain regime ($\alpha_p^2$) of the pump for the case when the idler beams are perfectly aligned. The violation in this regime is suppressed by misaligning the idler beams. {The explicit dependence of joint probabilities on $t$ are expressed as follows:
\begin{subequations} \label{subeqs3}
	\begin{align}
 P(0,1|1,2)&= \mean{ \Pi^{\rm off}_{\rm post} \Pi^{\rm on}_{i_{1,2}}}=\dfrac{1}{3}(1+t^2),\\
	P(0,1|2,3)&=\mean{ \Pi^{\rm off}_{i_{1,2}} \Pi^{\rm on}_{s_3}}=\dfrac{1}{3},\\
	P(0,1|3,4)&=\mean{ \Pi^{\rm off}_{s_3} \Pi^{\rm on}_{i_1}}=\dfrac{1}{3},\\
 P(0, 1|4, 5)&= \mean{ \Pi^{\rm off}_{i_1}\Pi^{\rm on}_{i_{2,3}} }=\frac{1}{3}(1+t^2), \\
	P(0,1|5,1)&=\mean{ \Pi^{\rm off}_{i_{2,3}}, \Pi^{\rm on}_{\rm post}}=\dfrac{1}{2}(\dfrac{1}{3}-\dfrac{t^2}{9}).
	\end{align}
\end{subequations}
The expectation values are obtained from Eq. (\ref{eq9}) conditioned on the occurrence of SPDC. Thus,
the value of the contextuality witness as a function of $t$ is
\begin{equation}
\kappa=\dfrac{3}{2}+\dfrac{11t^2}{18},
\end{equation}
which produces the yellow curve ($\alpha_{p}^{2}$) in Fig. \ref{fig6}\aa{(b)}. Note that Eq. (\ref{eq9}) reduces to Eq. (\ref{eq20}) upon photon pair generation and in the case of perfect alignment.
The violation disappears for $t\leq \sqrt{\tfrac{9}{11}}\simeq 0.904$. The general analytic expression of $\kappa$ reproducing the other curves by including higher-order approximation to take into account the high-gain pump is shown in Appendix \ref{ details of KCBS}. 

In the high-gain pump, the signals and idlers are generated in every three NLs and therefore all detectors should click. In other words, in the high gain, even though the induced coherence remains, the input state approaches the classical limit of light. Therefore the sum of five joint probabilities is smaller than the maximum limit for NCHV theory, and for higher power of alpha tends to be zero.
Figure \ref{fig6} shows the result of calculating the Simple Form of KCBS inequality for different regimes. Figure \ref{fig6}\aa{(b)} shows the KCBS inequality for amplitude transmissivity $t$ of beam splitters between down-conversion crystals. For the low-gain regime, When amplitude transmissivity increases and path identity is completed, the KCBS inequality is violated.

{Classical model of SPDC is sufficient to explain many features that appear in ZWM setup, such as quantum imaging, involving Laser light, whether being in the low- or high-gain regime, interacting with nonlinear media \cite{ClassicalmodelBoyd}. Here, we propose a counter-example in which a classical model, or more general, noncontextual hidden variable description,  cannot reproduce the observation. The dramatic effect of correlated detectors at the low-gain regime turns the process into a quantum behavior that violates a classical inequality. The example is reminiscent of heralded single-photon generation with weak enough laser pumping in an optical nonlinear medium; The detector's click on one beam projects the state of the other beam onto a single-photon Fock state, i.e., resulting in a nonclassical state. }

Violation of the noncontextual hidden variable model is argued to be associated with the nonexistence of positive-valued quasiprobability representation for preparation and measurements simultaneously \cite{Negativity}.

The vacuum projection at mode $a$ is a Gaussian and therefore is represented by positive-valued Wigner distribution over phase space, $\xi_{\proj0}(\lambda_{a})$.
The linear transformations also preserve the Gaussianity of the state. 
The probability of the projection onto the vacuum for a given state would read, $P(0)=\int d\lambda_a W_\rho (\lambda_a)\xi_{\proj0}(\lambda_{a})$ in which $W_\rho (\lambda_a)$ is the Wigner function associated with the prepared state $\rho$.
The Wigner representation  joint probabilities of the adopted measurements in our scheme may read as
\begin{align}
    P(0,1|i,j)& \\
    =\int d\lambda_a d\lambda_b  &W_{\rho_{is}}(\lambda_a, \lambda_b) \xi_{\proj0}(\lambda_a, i) \big(1- \xi_{\proj0}(\lambda_b, j)\big).
    \nonumber
\end{align}
Since the Vacuum projector and its linear transformations are Gaussian, its  Wigner representation is positive-valued. Therefore,
the violation should be attributed to the Wigner negativity of the input state of signal and idler beams, $\rho_{is}$. A sufficient amount of the Wigner function negativity is necessary to achieve the violation and the negativity of the input state is maximal in the single-photon regime, Eq. (\ref{eq17}).

\section{conclusion} 
\label{sec conclotion}
This paper presents an alternative way of revealing the nonclassical property of the ZWM setting by violating the adopted non-contextuality inequality.  The measurement scheme employs the correlations between various pairs of detectors, the data which go beyond interference information. The test based on the proposed scheme in this study is feasible within the current state-of-the-art experiment facilitates used in standard ZWM setup. The realization of hardy-like contextuality in the extended version of the ZWM setup, pumped by weak laser light, provides a shred of unambiguous evidence on the quantumness of induced coherence without stimulated emission such that it cannot be emulated by classical resources. {Furthermore, the proposal put forward a versatile platform for rigorous tests of contextuality involving sequential measurements with photons.}

Measuring interference visibility and coincidence counts confirm the occurrence of induced coherence without induced emission conditioned on restrictive assumptions, such as single-photon input state, while the violation of the NCHV model provides sufficient evidence for induced coherence without induced emission. The theoretical analysis presented in this article has broad implications in the fundamental characterization of SPDC processes and their classical limit. Furthermore, the adopted model is capable of implementing spatial and temporal correlations and therefore extending the applicability of our setup for implementing a simultaneous test of nonlocality on signal and idler photon pairs and single-particle contextuality on the signal photon. Therefore, our work enables experiments that combine nonlocality and contextuality and analyze the interplay between them \cite{Bell-KCBS}. An interesting future direction is to extend the study of the induced coherence to other tests regarding other approaches to nonclassicality which can be also performed with two-dimensional systems \cite{Catani, wheeler}.

\begin{acknowledgments}
We warmly thank Amin Babazadeh for the useful discussions on the possible implementations of the proposal.
\end{acknowledgments}

\appendix
\section{Appendix}
\subsection{The unitary operator of  SPDC}
\label{U for SPDC}
The unitary time evolution operator for SPDC is
\begin{equation}
\hat{U}_{\operatorname{DC}_j}(t)={\rm exp}\Big[\dfrac{1}{i\hbar}\int_{0}^{\tau} H_{j}(t)dt\Big].
\end{equation}
$\tau$ is the interaction time, usually the time taken by the pump to travel the crystal's length. The exponential expansion is rewritten as follows:
\begin{equation}
\hat{U}_{\operatorname{DC}_j}(t)=1+ \dfrac{1}{i\hbar}\int_{0}^{\tau} H_{j}(t) dt + \dfrac{1}{2!} \Big[\dfrac{1}{i\hbar}\int_{0}^{\tau} H_{j}(t) dt\Big]^{2}+\cdots .\label{eqint-H}
\end{equation}

The following mathematical relationship is used to solve the integral:
\begin{equation}
\int {\rm exp}(ixz)dz=\dfrac{1-{\rm exp}(ix)}{ix}={\rm exp}(\dfrac{ix}{2}) {\rm sinc} (\dfrac{x}{2} ).
\end{equation}
For phase match condition $\omega_{p}=\omega_{s}+\omega_{i}$, therefore expressions of ${\rm exp}(\bigtriangleup\omega)$ and ${\rm sinc}(\bigtriangleup\omega) $ lead to 1. Therefore, the Eq. (\ref{eqint-H}) is written as follows
\begin{equation}
\hat{U}_{\operatorname{DC}_j}(t)=1+g\hat{a}_{{p}_{j}} \hat{a}^{\dagger}_{{s}_{j}}\hat{a}^{\dagger}_{{i}_{j}} + \dfrac{g^{2}}{2}(\hat{a}_{{p}_{j}} \hat{a}^{\dagger}_{{s}_{j}}\hat{a}^{\dagger}_{{i}_{j}})^{2}+\cdots ,
\end{equation}
where $ g= \dfrac{g\prime\tau}{i\hbar} {\rm exp}(\dfrac{i\bigtriangleup\omega\tau}{2}) {\rm sinc} (\dfrac{\bigtriangleup\omega\tau}{2} )$, and $g\prime$ contains the same order of $g$. Although their explicit forms are not necessary for the purpose of our discussion.
\subsection{ Explicit form of Tritter unitaries}
\label{tritter}
The reader might be interested to see the explicit form of the tritter unitary which is a combination of two beam splitters and a mirror \cite{multiportBS}.
Imagine a single-photon $\ket{\psi_{\rm in}}=\ket{1_p}$. Such a photon can be routed to one of the three NLs using the first tritter 
$U_i=U_{\operatorname{BS}_2}U_{\operatorname{BS}_1}$
 as
\begin{equation}
U_i=\frac{1}{\sqrt{3}}\begin{bmatrix}
1 & -\sqrt{2} & 0 \\
1 & \frac{1}{\sqrt{2}} & -\frac{\sqrt{3}}{\sqrt{2}} \\
1 & \frac{1}{\sqrt{2}} & \frac{\sqrt{3}}{\sqrt{2}} 
\end{bmatrix} \label{1a}.
\end{equation}
The matrix $U_{i}$ is a combination of two rotation matrices around the axis of $z$ and $x$ with transmission coefficients of $\tfrac{1}{\sqrt{3}}$ and $\tfrac{1}{\sqrt{2}}$ for $\operatorname{BS1}$ and $\operatorname{BS2}$, respectively, and a mirror $\mathrm{M1}$.
Similar to the $U_i$, the second tritter $U_f=U_{\operatorname{PS}}U_{\operatorname{BS}_4}U_{\operatorname{BS}_3}$ consists of two beam splitters ($\operatorname{BS3}$ with a transmission coefficient of $\tfrac{1}{\sqrt{3}}$ and $\operatorname{BS4}$ with a transmission coefficient of $\tfrac{1}{\sqrt{2}}$), along with a phase shifter in path $\mathrm s_2$ and a mirror M2 (Fig. \ref{fig2})
\begin{equation}
U_f= \frac{1}{\sqrt{3}}\begin{bmatrix}
1 & -\sqrt{2} & 0 \\
-1 & -\frac{1}{\sqrt{2}} & \frac{\sqrt{3}}{\sqrt{2}} \\
1 & \frac{1}{\sqrt{2}} & \frac{\sqrt{3}}{\sqrt{2}} 
\end{bmatrix} \label{2a}.
\end{equation}

\subsection{Pre- and post-selected paradoxes}
\label{sec pre and post}
The study of quantum systems that are both pre-and post-selected (PPS) was initiated by Aharonov, Bergmann, and Lebowitz (ABL) in 1964 \cite{PPS}. A prototype example is known as the “three-box paradox” exhibiting a peculiar contextual behavior. Imagine a quantum particle that can be in one of three boxes, $k=1,2,3$. The state of the particle being in box $k$ is denoted by $\ket{k}$. Suppose the particle's state is prepared in a superposition $\ket{\psi_{\rm pre}}=\frac{1}{\sqrt{3}}(\ket{1}+\ket{2}+\ket{3})$, and project it later onto a non-orthogonal state $\ket{\psi_{\rm post}}=\frac{1}{\sqrt{3}}(\ket{1}-\ket{2}+\ket{3})$. One can think of the postselection as a dichotomic measurement. Hence, it is described by
$\{\Pi_{\rm post}=\proj{\psi_{\rm post}}, \mathbbm 1-\Pi_{\rm post} \}$ that clicks on positive results, obtaining $\ket{\psi_{\rm post}}$ with success probability of $1/9$ and discarding the negative result. In the PPS scenario, $\ket{\psi_{\rm pre}}$ and $\ket{\psi_{\rm post}}$ are called pre- and post-selected states, respectively. 

This innocent-looking example leads to counter-intuitive predictions when certain intermediate dichotomic measurements, $M_k$, are introduced on the pre-and post-selected system. The PPS setting, therefore, can be viewed as a sequence of two measurements with associated joint probabilities,
\begin{align}
    p_{\psi_{\rm pre}}(k, \psi_{\rm post}) &=\bra{\psi_{\rm pre}}  \Pi_k\Pi_{\rm post}\Pi_k \ket{\psi_{\rm pre}} \nonumber \\
    &= |\mean{\psi_{\rm post}|\Pi_{k}|\psi_{\rm pre}}|^2.
    \label{Pjoint}
\end{align}
Here the intermediate measurement $M_k$ is described by the following set of projectors \cite{PPSandcontextuality} $\{\Pi_{k}=\proj{k}, \Pi_{\bar k}=\mathbbm 1-\proj{k} \}$ and $k=1,3$, whereas, $\Pi_{\rm post}$ deemed the second measurement's postselection projector. 

State \eqref{eq20} gives the joint probability
\begin{equation}
    P( 1, \psi_{\rm post})= \mean{\Pi^{\rm on}_{i_{1}} \Pi^{\rm on}_{\rm post} \Pi^{\rm on}_{i_{1}}}  =\frac{1}{3}.
     \label{eq21}
\end{equation}

By a straightforward application of Bayes’ theorem to the joint probability one can derive the probability of finding the particle in box $k$ conditioned on the pre- and post-selected states \cite{PPS,PPSandcontextuality,nonclassicalinPPS,Spekkens,time-symmetric},
\begin{equation}
p(k|\psi_{\rm pre}, \psi_{\rm post})=\dfrac{|\mean{\psi_{\rm post}|\Pi_{k}|\psi_{\rm pre}}|^{2}}{\sum_{k}|\mean{\psi_{\rm post}| \Pi_{k} |\psi_{\rm pre}}|^{2}}. \label{eq15}
\end{equation}
The sticking fact about the above expression is that it implies a unit probability of finding the particle regardless of which measurement, $M_1$ or $M_3$ is chosen for such pre- and postselection. From the classical point of view, the sum of the probability of the mutually exclusive event is bounded by one, that is,
\begin{equation}
p(1|\psi_{\rm pre}, \psi_{\rm post})+p(3|\psi_{\rm pre}, \psi_{\rm post})\leq 1,
\label{eq 16}
\end{equation}
which is violated in this quantum experiment. Since $p(k|\psi_{\rm pre}, \psi_{\rm post})$ is not predictive but rather retrodictive, that is the intermediate measurement outcome is conditioned on the future (post-selected) outcome as well, therefore inequality of Eq. (\ref{eq 16}) cannot be directly tested.  

Let us clarify a point here. As far as the joint probabilities are concerned the final measurement does not need to realize the post-measurement state $\ket{\psi_{\rm post}}$, called the L\"uder instrument. As a result, it is sufficient to only read out the outcome from the detector $D_s$.

{
\subsection{The analytic result of $\kappa$ }
\label{ details of KCBS}
Contextuality witness measure $\kappa$ reproducing the $k$-th order of approximation illustrated in Fig. \ref{fig6} is given by 
\begin{align}
   \kappa= & 
    \sum _{n=1}^k c_n \kappa^{(n)},
\end{align}
where
\begin{subequations}
\begin{align}
    \kappa^{(1)}=& \left(\frac{11 t^{2}}{18} + \frac{3}{2}\right), \\
    \kappa^{(2)}=& \left(- \frac{61 t^{4}}{324} + \frac{4 t^{2}}{9} + \frac{37}{36}\right), \\
   \kappa^{(3)}=& \left(\frac{13 t^{4}}{243} + \frac{32 t^{2}}{243}  + \frac{1}{3}\right), \\
   \kappa^{(4)}=&\left(\frac{385 t^{4}}{8748} + \frac{263 t^{2}}{4374} + \frac{493}{8748}\right).
\end{align}
\end{subequations}
The expansion coefficients are
\begin{equation}
    c_n=\frac{\alpha^{2n}g^{2n}}{\sum_{n=1}^k \alpha^{2n}g^{2n}}.
\end{equation}
}
\subsection{Symbolic description language}
\label{symbol}
Symbolic algebra is a powerful tool for representing quantum states. This mathematical approach allows for the description of quantum states in a concise and precise manner. A single-photon state,  for example, can be represented using the following notation:
\begin{equation}
    \mathrm{a}[1, d_1, ... , d_n],
\end{equation}
where $\mathrm{a}$ denotes the path traversed by the photon and $d_i$ represents the properties of different degrees of freedom, such as polarization, orbital angular momentum, and so on. It is also possible to generate the unitary operator for each experimental element by representing the creation and annihilation operators through symbolic algebra.
\begin{subequations}
\begin{align}
\hat a^{\dagger}[\psi_{\text{in}}, \mathrm{a}]&=\psi_{\text{out}} \Leftarrow\{\mathrm{a}[n] \rightarrow \sqrt{n+1} \mathrm{a}[n+1],\label{41a} \\
\hat a[\psi_{\text{in}}, \mathrm{a}]&=\psi_{\text{out}}\Leftarrow\{\mathrm{a}[n] \rightarrow \sqrt{n} \mathrm{a}[n-1].\label{41b}
\end{align}
\end{subequations} 
The symbol $\Leftarrow$ indicates a symbolic replacement and $n$ is the number of photons. To simplify our analysis, it is assumed that all photon degrees of freedom are the same. As an example, after passing through a 50:50 symmetric beam splitter ($r = t = \tfrac{1}{\sqrt{2}}$), the symbolic transformation of a single-photon $\mathrm{a}[1]\mathrm{b}[0]$ $(\hat a^{\dagger}\mathrm{a}[0]\mathrm{b}[0])$ can be defined as follows:
\begin{align}
\mathrm{\operatorname{BS}}[\psi_{\text{in}},\mathrm{a},\mathrm{b}]&=\psi_{\text{out}}\Leftarrow\{\mathrm{a}[1]\mathrm{b}[0]\notag\\
&\to \frac{1}{\sqrt{2}} (\mathrm{a}[1]\mathrm{b}[0]+\mathrm{a}[0]\mathrm{b}[1]),
\end{align}
here $\mathrm{a}$ and $\mathrm{b}$ are paths of photon. $r$ and $t$ denote the amplitude of reflection and transmittance of the beam splitter, respectively. The unitary transformation of the BS according to the creation operator is $\tfrac{1}{\sqrt{2}} (\hat a^{\dagger}+\hat b^{\dagger})$ \cite{gerry}.  The output state shows that a single-photon incident at one of the input ports of the beam splitter ($\mathrm{a}$), the other port containing only the vacuum ($\mathrm{b}$) will be either transmitted or reflected with equal probability.
The symbolic algebra to perform calculations facilitates the effortless addition of new elements and degrees of freedom to quantum optics experiments. Moreover, this method enhances the comprehensibility of the results for human interpretation. Therefore, the ZWM experiment can be expressed as follows: 
\begin{align}
\operatorname{ZWM}&[\psi_{\text{in}}, p_1,  p_2, s_1, s_2, i_1, i_2]\notag\\
=& \operatorname{BS}[\operatorname{PI}[\operatorname{NL}[\operatorname{O}[\operatorname{NL}[\operatorname{BS}[\psi_\text{in}, p_1,p_2], p_1, s_1, i_1], i_1],\notag\\
& p_2, s_2 , i_2],i_1, i_2], s_1, s_2].
\end{align}
\bibliography{refs}
\end{document}